\documentclass[preprint2]{aastex}

\newcommand{\gapprox}{\stackrel{>}{\scriptstyle\sim}}

\slugcomment{Accepted by Ap. J. Letters in October 2001.}

\shorttitle{SGT for the Vela Pulsar}
\shortauthors{Cairns et al.}

\begin{document}


\title{Intrinsic Variability of the Vela Pulsar: Lognormal 
Statistics and Theoretical Implications}


\author{Iver H. Cairns, S. Johnston, and P. Das}

\affil{School of Physics, University of  Sydney, Sydney,  
Australia.}


\begin{abstract}
Individual pulses from pulsars have 
intensity-phase profiles that differ widely from pulse to pulse, 
from the average profile, and from phase to phase within a pulse. 
Widely accepted explanations for pulsar radio emission and its time 
variability do not exist. Here, by analysing data near the peak 
of the Vela pulsar's average profile, we show that Vela's variability 
corresponds to lognormal field statistics, consistent with the 
prediction of stochastic growth theory (SGT) for a purely linear 
system close to marginal stability. Vela's variability is therefore 
a direct manifestation of an SGT state and the field statistics 
constrain the emission mechanism to be linear (either direct or indirect), 
ruling out nonlinear mechanisms like wave collapse. Field statistics 
are thus a powerful, potentially widely applicable tool for understanding 
variability and constraining mechanisms and source characteristics 
of coherent astrophysical and space emissions. 
\end{abstract}




\section{Introduction}

   Pulsars are highly magnetized neutron 
stars whose rotation causes highly nonthermal beams of radiation to 
be swept across the Earth \citep{man77}, similar to the periodic 
viewing of a lighthouse 
beam. Most likely the radio emission is produced over the star's (magnetic) 
polar caps \citep{man77,melrose1996,asseo1996}. Since their 
discovery in 1967 it has been recognized 
that only suitably long time averaging leads to a stable intensity profile 
versus pulsar phase.  While this average profile is unique, 
individual pulses vary widely in 
intensity, often by a factor of $5$ or more, from one phase to 
another in a given pulse and from one pulse to the next at a given phase, 
as in Fig. 1.  
This variability \citep{man77,hankins1996} includes 
phenomena known as drifting sub-pulses \citep{drakecraft1968},  
microstructure \citep{craftetal1968}, giant pulses \citep{cognardetal1996} and 
giant micropulses \citep{johnstonetal2001}. Sub-pulses 
are features which drift in time across the pulse window whereas microstructure 
are concentrated features superposed on a subpulse that are sometimes  
quasiperiodic. Giant pulses and micropulses are very rare pulses with fluxes 
$\gapprox 10$ times the average flux \citep{cognardetal1996,johnstonetal2001}. 
No accepted explanation exists for these forms of variability or, indeed, for 
the mechanism(s) producing pulsar radio emission
\citep{man77,melrose1996,asseo1996,hankins1996}. 

   The high brightness temperatures of pulsars 
require coherent emission processes such as plasma microinstabilities 
or nonlinear processes. Linear mechanisms include 
\citep{melrose1996,asseo1996,luomelrose1995}:(1) linear 
acceleration and maser 
curvature emission, in which electrons radiate coherently while 
accelerating in an oscillating large-scale field or on curved magnetic 
field lines, respectively, and (2) relativistic plasma emission, in which a 
streaming instability either directly generates escaping radiation 
near harmonics of the electron plasma frequency $f_{pe}$ or else drives 
localized (non-escaping) waves near $f_{pe}$ that are converted into 
escaping harmonic radiation by linear mode conversion or nonlinear processes. 
Nonlinear mechanisms can produce radiation by wave coalescence and 
scattering processes or as intense localized 
wavepackets, perhaps driven near $f_{pe}$ by a streaming instability, 
undergo modulational instabilities and strong turbulence wave collapse 
\citep{asseoetal1990,asseo1996,weatherall1998}. 
Since existing analyses suggest that many mechanisms are viable, in part 
due to large uncertainties in the plasma properties and location of the 
emitting regions (e.g., above the polar cap or near the light cylinder), new 
approaches are necessary. 
 
    Analyses of intensity scintillations and angular broadening, corresponding 
primarily to Fourier analyses of data, are standard for astrophysical and solar 
system radiation sources \citep{rickett1990}. In contrast, 
distributions of electric field strengths 
or intensities were rarely analyzed until recently, perhaps because  
their strong theoretical motivations and benefits were not clear before the 
advent of stochastic growth theory (SGT) \citep{r1992,retal1993,cr1999,cetal2000,rc2001} 
and other theories like self-organized criticality (SOC) 
\citep{baketal1987}. However, recent analyses of 7 different solar system wave 
phenomena  
show that all have well-defined field distributions that agree very well with 
the predictions of SGT \citep{retal1993,cr1999,cetal2000,rc2001}, 
resolving longstanding 
theoretical problems pertaining to the burstyness, widely varying fields, 
and persistence of the waves. Similarly, the giant pulses 
of some pulsars have power-law flux distributions \citep{cognardetal1996}, sometimes 
interpreted qualitatively in terms of SOC \citep{youngkenny1996}. With the 
advent of rapid time resolution, coherently de-dispersed data for Vela and other 
pulsars \citep{johnstonetal2001} the time is ripe for analyzing pulsar 
variability and its statistics in terms of SGT, SOC and other theories. 

    This paper directly addresses pulsar variability and emission mechanisms by 
analysing the radiation's statistics near the peak of the Vela pulsar's average profile 
and interpreting the results in terms of the theoretical predictions and 
formalism of SGT. 
After summarizing the predictions of SGT and other theories for wave growth (Section 2), 
Section 3 shows that Vela's 
intrinsic variability near the peak of the average pulse profile corresponds to 
lognormal statistics in the electric field (or intensity), not 
Gaussian or power-law statistics in the intensity. Pulse variability is thus  
a direct manifestation of an SGT state. The consistency 
with the SGT prediction then strongly constrains the emission mechanism 
and source plasma (Section 4), with nonlinear emission mechanisms being non-viable 
in the phase range analysed. Preliminary results at other phases of Vela's 
pulse profile and for other pulsars are then briefly discussed (Section 5). This analysis 
provides a first demonstration that radiation statistics for 
astrophysical sources are a powerful and potentially widely applicable 
tool for strongly constraining emission mechanisms and source plasmas.

\section{Theories for Wave Statistics}

   Wave-particle interactions are expected to drive natural plasmas towards 
marginal stability, where wave emission and damping (as well as total energy 
inflow and outflow) are balanced. SGT treats systems in which an unstable 
particle distribution interacts self-consistently with its driven waves in an 
inhomogeneous plasma background and evolves to a state in which (i) the 
particle distribution is close to time- and volume-averaged marginal stability but 
with stochastic fluctuations that (ii) cause the wave gain $G$ to be a stochastic 
variable \citep{r1992,retal1993,cr1999,cetal2000,rc2001}. 
Here $G(t)$ is related to the wave growth 
rate $\Gamma(t)$ by 
$G(t) = \int_{-\infty}^{t} dt' \Gamma(t')$ and to the time-varying wave electric 
field $E(t)$ and a reference field $E_{0}$ by $E^{2}(t) = E_{0}^{2} \exp G(t)$. 
Rewriting this time integral as a summation over fluctuations $\Delta G_{i} 
= \Gamma_{i} \Delta t_{i}$ then, provided only that sufficiently many 
fluctuations in $\Delta G_{i}$ occur in some characteristic time, the Central 
Limit Theorem requires that $G(t)$ is a Gaussian random variable irrespective 
of the detailed distribution of $\Delta G_{i}$. Hypotheses (i) and (ii) thus 
have simple and natural physical justifications. The hypothesized random walk 
in $G \propto \ln E$ then implies that 
the waves should be bursty and widely varying in amplitude, while the closeness 
to marginal stability implies that the waves and driving distribution should 
persist far from the latter's source. These characteristics 
are very attractive for pulsars, given the existence of intrinsic variability 
and the radiation's broad bandwidth (and so large radial extension of the 
source inferred therefrom), 
as well for many other astrophysical and space phenomena. 

   Due to $G$ being a Gaussian random variable, pure SGT predicts that the 
probability distributions of wave field and intensity are lognormal 
\citep{retal1993,cr1999,rc2001}; i.e., 
\begin{equation}
P(\log E) = (\sqrt{2\pi}\ \sigma)^{-1} 
\exp( - (\log E - \mu)^{2} / 2\sigma^{2} )\ , 
\end{equation} 
where $\log$ means to the base $10$, $\mu$ and $\sigma$ are the average 
and standard deviation of $\log E$, respectively,   
and $\int d({\rm log} E) P(\log E) = 1$. 
Nonlinear 3-wave processes active at high $E$ above a threshold $E_{c}$, which 
remove energy from the waves, reduce the $P(\log E)$ distribution below the 
prediction (1) near and above $E_{c}$ with known analytic form \citep{retal1993,rc2001}. 
Processes like wave collapse and modulational instability 
cause a power-law tail with $P(E) \propto E^{-\alpha}$, with $\alpha$ ranging from $4$ to $6$, 
to develop above $E_{c}$ \citep{r1997,rc2001}. Waves driven from thermal 
levels by an instability, and which 
retain memory of their thermal past, also develop a power-law tail, but usually 
with a smaller index \citep{cetal2000,rc2001}. Finally, SOC should produce a 
power-law distribution with index close to $-1$ \citep{baketal1987} and the usual model 
for wave growth in plasmas (uniform secular growth with constant growth rate) should 
produce a uniform distribution at fields below $E_{c}$ \citep{retal1993,cr1999,rc2001}. 
In contrast, scattering by density turbulence or radiation from multiple
incoherently 
superposed sources is expected to produce Gaussian 
intensity distributions \citep{rickett1990}. Rigorous testing of theories 
for wave growth is thus possible using the observed field statistics, as 
already demonstrated in multiple space contexts referenced above. 


\section{Statistics of Vela's intrinsic variability}  

   The data set consists of 20085 contiguous pulses (30 minutes) of the 
Vela pulsar, measured at 1413 MHz by the Parkes radio telescope and processed 
using coherent dedispersion and other techniques by 
\citet{johnstonetal2001}. There 
are 2048 phase (time) bins per pulse period, each of $44\ \mu$s length (comparable to
the scatter broadening time). Vela's average intensity (over many pulses) is restored by 
adding $I_{0} = 1250$ mJy to each sample. Fig. 1 shows the average pulse profile for 
relevant phase bins in mJy, together with $3$ superposed pulses which illustrate 
the variability. Note that the noise level is very low compared with earlier analyses, allowing 
detailed investigation of the intrinsic field statistics. 

   Analysis of data in the off-pulse phase bins lead to Gaussian statistics   
in the intensity $I$, as expected for instrumental and background noise. For 
instance,  fitting the $P(I)$ distribution for phases 391-399 to a Gaussian, 
by using the Amoeba algorithm 
to minimize $\chi^{2}$ \citep{pressetal1986}, yields $\langle I \rangle = 1215$ mJy (agreeing 
with $I_{0}$ to within less than the $100$ mJy  bin width), 
$\sigma_{I} = 1420$ mJy, $\chi^{2} = 66$ for $N_{DF} = 46$ degrees of freedom, and a 
significance probability $P(\chi^{2}) = 0.03$. (Fitting is restricted to intensity 
bins with $\ge 100$ pulse samples). This fit has good statistical significance. 

   Fig. 2 shows the $P(I)$ distribution and its best Gaussian fit for phase bin 
490, close to the 
peak in Vela's average profile.  The fit clearly fails at both low and high $I$, 
entirely missing the long tail at large $I$, as confirmed by it  
having $\chi^{2} = 301$ for $N_{DF} = 53$ and 
$P(\chi^{2}) < 10^{-36}$. The variability at this phase is thus not described by Gaussian 
intensity statistics. 

   In contrast, defining the electric field $E = I({\rm mJy})^{1/2}$, 
Fig. 3 shows that the $P(\log E)$ distribution for 
phase 490 is well fitted by the SGT prediction (1): for bins with $\ge 100$ 
pulse samples and $E \ge 10^{2}$ units (intensities above $10^4$ mJy, which 
is $6\sigma_{I}$ above the noise), $\mu = 2.3$, $\sigma = 0.096$, $\chi^{2} = 27$ for 
$N_{DF} = 19$, and 
$P(\chi^{2}) = 0.12$. The Kolmogorov-Smirnov test \citep{pressetal1986} 
yields a significance 
probability of $47 \%$. This fit is strongly statistically significant, 
clearly demonstrating that pulsar variability at this phase is 
lognormally distributed and quantitatively consistent with the 
theoretical form predicted by simple SGT. The fit matches the data 
well even outside the fitted range of fields (dotted line), although 
the effects of the noise background become increasingly evident at 
fields $\le 80$ units. 

   Results similar to Fig. 3 are found for phases 485 - 540, for 
which the average pulsar intensity is well above the noise level, although 
the statistical significance varies. Rather than showing more results for 
individual phases, Fig. 4 shows the $P(X)$ distribution observed for phases 485 - 500 
simultaneously, where $X = (\log E - \mu(\phi) ) / \sigma(\phi)$ 
is the field variable resulting from detrending variations in 
$\mu$ and $\sigma$ with phase $\phi$. Comparison with (1) 
shows that SGT predicts the $P(X)$ distribution to be Gaussian with zero  
mean and unit standard deviation (solid curve) \citep{cr1999}. The agreement is 
very good, with the Kolmogorov-Smirnov test yielding a significance probability 
of $0.1\%$. 

\section{Theoretical Implications for Vela}

   For phases 485 - 540, where Vela's average pulse profile is well 
above the noise, the field distributions do not have 
power-law tails or nonlinear cutoffs. Instead, the data 
have lognormal statistics and the variability is a direct manifestation 
of a simple 
SGT state, with no evidence for nonlinear processes, SOC, or uniform 
secular growth. This absence of a power-law tail or cutoff in the 
$P(\log E)$ distributions 
for these phases rules out pulsar 
emission mechanisms based on nonlinear processes 
\citep{asseoetal1990,asseo1996,weatherall1998} such as wave 
collapse, modulational instability, and three-wave processes. Instead, the 
observed consistency with 
simple (linear) SGT means that only linear emission mechanisms are viable, 
meaning that a plasma instability in a SGT state either directly generates 
the radiation or else generates non-escaping waves that are tranformed 
into escaping radiation by linear processes (e.g., mode conversion) alone.  

   From the definitions of $\mu$ and $\sigma$ and the intensity decreasing 
with distance $R$ as $R^{-2}$, it is easy to show that $\sigma(R)$ is  
constant and that $\mu(R) = \mu(R_{0}) - \log(R/R_{0})$, 
where $\mu(R_{0})$ is the value at the source's edge ($R = R_{0}$). 
Taking the values $\mu = 2.0$ and $\sigma = 0.1$ to be representative 
of these phases, the distance $R = 350$ pc for Vela, and the 
value $R_{0} = 30$ m, yields $\mu(R_{0}) \approx 20$. 
The value $R_{0} = 30$ m results from assuming that the overall 
source is annular, with radius equal to the neutron star radius $\approx 
10$ km, and dividing by the $2048$ phase bins used for Vela. Accordingly, the ratio 
$\mu_{0}/\sigma \approx 200$ in the source. The values $\mu_{0}$ and 
$\sigma$ will constrain future theoretical models for why SGT applies. 

\section{ Discussion and Conclusions}

   The foregoing analyses are the first applications of SGT to propagating 
EM radiation and, simultaneously, to extra-solar system sources. Their 
success implies that radiation statistics are an underappreciated and 
potentially very powerful tool in astrophysics (and space physics), and 
suggests that SGT may well be widely applicable to coherent astrophysical 
sources. As to whether the Vela results are representative of other 
pulsars, analyses are ongoing. Our results 
to date for pulsar PSR 1641-45, see also \citet{johnstonromani2001}, suggest that 
the variability 
near the peak of the average profile also corresponds to lognormal 
statistics, thereby being consistent with SGT and the Vela results above.

   Of course, SGT is not likely applicable to all sources or indeed to all 
components of pulsar emissions. For 
instance, Jovian ``S bursts'' have a power-law  flux 
distribution with index $2.0 \pm 0.5$ \citep{queinneczarka2001} and the 
peak flux distribution of 
solar microwave spikes can be fitted with an exponential or perhaps a lognormal form 
\citep{islikerbenz2001}. Moreover, this richness in possible wave 
statistics also appears in phase bins away from the peak in Vela's 
average profile and for pulsars with giant pulses, where the observed 
$P(\log E)$ distributions are often approximately power-law. Detailed interpretations 
will be described in detail elsewhere. 
For now, we mention only that indices $\approx 4.5 \pm 1.0$ are likely too high for SOC but 
are instead probably due to either driven thermal 
waves \citep{cetal2000} and/or due to strongly nonlinear processes like modulational 
instability and wave 
collapse \citep{r1997,rc2001}. The latter idea complements earlier suggestions
\citep{asseoetal1990,weatherall1998} and appears particularly attractive 
for giant pulses and giant micropulses. 

  In conclusion, analysis of rapidly-sampled, coherently de-dispersed data near 
the peak of the Vela pulsar's average intensity-phase profile show that the field 
statistics are lognormal and quantitatively consistent with SGT's prediction 
for a purely linear system near marginal stability. The variability is thus 
a direct manifestation of an SGT state and only linear emission mechanisms (either 
direct or indirect) are viable. Observations for other pulsars and at other phases 
for Vela yield both similar and different results, hinting at a possible richness 
of wave statistics and emission mechanisms. Analysis of field statistics is thus a powerful 
tool for understanding source variability and constraining the emission 
mechanisms and source characteristics that may be widely useful for coherent 
astrophysical and solar system radio emissions, as already found for plasma 
waves in space.



\acknowledgments

 The Australian Research Council and University of Sydney 
 funded this work. We thank P.A. 
Robinson, Q. Luo, D.B. Melrose, B.J. Rickett, and R. Romani for 
helpful discussions.





\clearpage

\begin{figure}
\epsscale{0.8}
\plotone{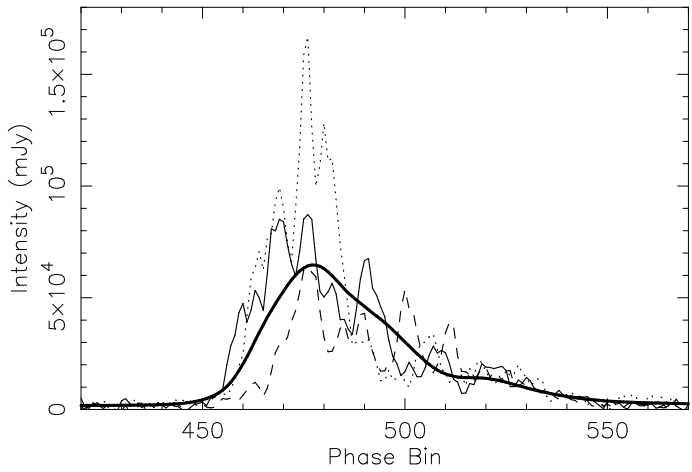}
\caption{Average intensity-phase profile of the Vela pulsar for the dataset (thick solid line) 
together with three superposed individual pulses (other lines).}
\end{figure}

\begin{figure}
\plotone{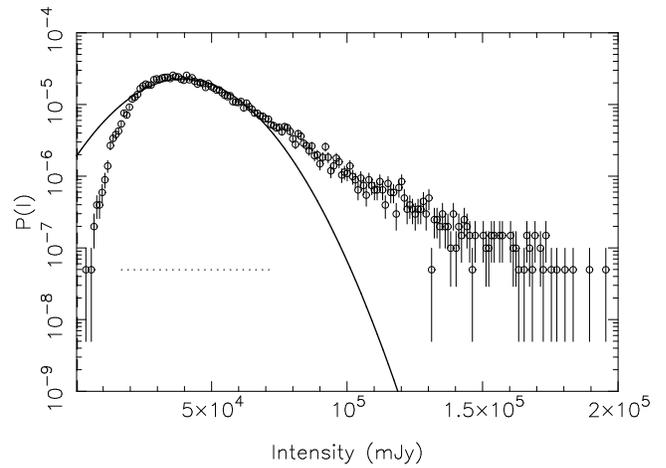}
\caption{The observed distribution $P(I)$ at phase 490, formed by binning the data linearly in the 
intensity and normalizing by $\int dI P(I) = 1$, is shown with open circles and 
$\pm \sqrt{N}$ error bars. The solid line shows the best-fit Gaussian, calculated for the 
domain shown by the dashed line.}   
\end{figure}

\begin{figure}
\plotone{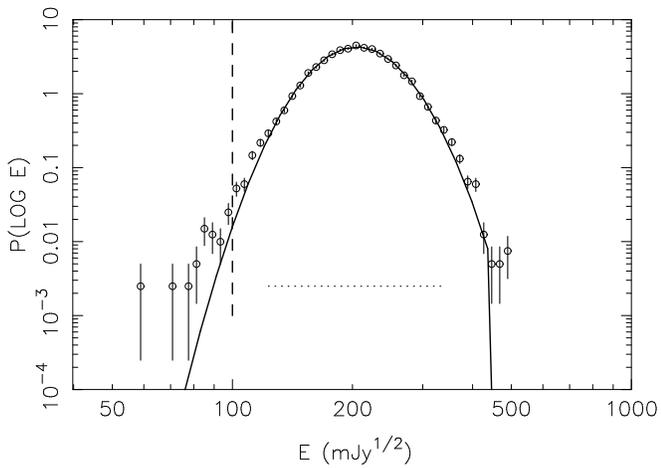}
\caption{The observed distribution $P(\log E)$ at phase 490 (open circles and $\pm \sqrt{N}$ error bars), 
formed by binning the data linearly in $\log E$ and normalizing, is compared with the 
prediction (1) for lognormal field statistics. The symbols and dashed line are as in Figure 2. }  
\end{figure}

\begin{figure}
\epsscale{1.0}
\plotone{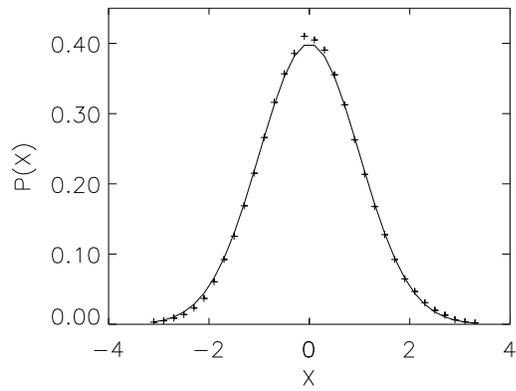}
\caption{The distribution $P(X)$ for all data in phases 
485 -- 500 (crosses), inclusive, is compared with the SGT prediction (solid line). 
See text for details. }   
\end{figure}

\end{document}